\documentstyle[12pt]{article}
\global\arraycolsep=2pt
 \newcommand{\be}{\begin{eqnarray}}
\newcommand{\ee}{\end{eqnarray}}
\newcommand{\la}{\langle}
\newcommand{\ra}{\rangle}
\input epsf
\begin{document}
\begin{titlepage}
\begin{flushright}
 hep-ph/9604314\\
April 1996
\end{flushright}
\vspace{0.3cm}
\begin{center}
\Large\bf
Strange and Charmed Quarks in the Nucleon.
\end{center}
\vspace {0.3cm}

 \begin{center}    {\bf Ariel R. Zhitnitsky}\footnote{
  e-mail address:arz@physics.ubc.ca }
 \end{center}
\begin{center}
{\it Physics Department, University of British Columbia,
6224 Agricultural  Road, Vancouver, BC V6T 1Z1, Canada}
\end{center}
 \begin{abstract}
We discuss the general method of the calculation of the nucleon
matrix elements of an operator associated with nonvalence quarks.
The method is based on the QCD sum rules and low energy theorems.
As an application of these considerations, we calculate
the strange quark matrix element as well as the momentum
distribution of the strangeness in the nucleon.
We also calculate the  singlet axial constant associated with $\eta'$ meson
as well as an   axial constant associated with heavy quarks.

\end{abstract}
\begin{center}
{\it Talk given at the Workshop on the Structure of the $\eta'$ Meson.\\
NMSU and CEBAF, Las Cruces, New Mexico, March 8,1996. \\ }
\end{center}

\end{titlepage}
\vskip 0.3cm
\noindent

\setcounter{page}{1}

\section{Introduction}
 For a long time it was widely believed that the admixture of the pairs of
strange quarks  in the nucleons is small. The main justification
of  this  picture was the constituent quark model where
there is no room for strange quark in the nucleon.
  It has been known for a while that  this picture is not quite true:
In scalar and pseudoscalar   channels
one can expect a noticeable deviation from this naive prediction.
This is because,
these channels are very unique in a sense that they are
tightly connected to the QCD-vacuum fluctuations  with $0^{+},0^{-}$
singlet  quantum numbers.
Manifestation of the uniqueness
can be seen, in particular, in the existence of the axial anomaly ($0^{-}$
channel) and the trace anomaly ($0^+$ channel).
Nontrivial QCD vacuum structure tells us that
one could expect some unusual properties when we   deal
with those quantum numbers.

As we now  know , this is indeed the case.
In particular, we know that
  the strange quark matrix
element $\la N|\bar{s}s|N\ra$ does not vanish
 and has the same order of magnitude as $\la N|\bar{d}d|N\ra$.
This information can be obtained from the analysis of the so-called
$\sigma$ term \cite{Cheng},\cite{Koch}. Similarly the analysis of   the `` proton
spin crisis''  essentially teaches us that the spin which is carried by the
strange quark in the nucleon  is not small
as naively one could expect, see e.g. the recent review
\cite{Ellis}.

 Another phenomenological manifestation of
the same kind is the
very old observation that in the scalar and pseudoscalar
channels the Zweig rule is badly broken and there is substantial
admixture of $s$ quarks in the scalar mesons $f_0 (980)$ (was
$S^*$), $a_0 (980)$ (was $\delta$),
$f_0 (1300)$ (was
$\epsilon$), as well as in the pseudoscalar  mesons $\eta$ and $\eta'$.
At the same time, in the vector channel the
Zweig rule works well. Phenomenologically it is evident in e.g.
the smallness of the $\phi-\omega$ mixing. In terms of   QCD
such a smallness corresponds to the numerical suppression
of the nondiagonal correlation function
$\int dx \la 0|T\{\bar{s}\gamma_{\mu}s(x), \bar{u}\gamma_{\nu}u(0)\}|0\ra$
in comparison with the diagonal one
$\int dx \la 0|T\{\bar{u}\gamma_{\mu}u(x), \bar{u}\gamma_{\nu}u(0)\}|0\ra$.
In the scalar and pseudoscalar channels diagonal and non-diagonal channels
have the same order of magnitude. We believe that analysis
of such kind of the correlation functions  is an appropriate
method for a QCD- based explanation of the  unusual
hadronic properties mentioned above.

In this talk we  present some general methods and ideas
for  the analysis of the nucleon matrix elements from a non-valence operator.
The ideology and methods (unitarity,
dispersion relations, duality, low-energy theorems) we use  are motivated
by QCD sum rules. However we do not use the QCD sum rules in the common sense.
Instead, we reduce one complicated problem
(the calculation of non-valence nucleon matrix elements)
to another one (the behavior of some vacuum correlation functions
at low momentum transfer).
One could think that such a reducing of one problem to another one
(may be even more complicated) does not improve  our  understanding
 of the phenomenon. However, this is not quite true:
The analysis of the vacuum correlation functions  with vacuum quantum numbers,
 certainly, is a very difficult problem. However
some nonperturbative information based on the low energy theorems
is available for such a correlation function.
   This gives some chance to estimate some interesting quantities.
\section{Strangeness in the nucleon, $0^+$ channel.}
\subsection{First estimations}
We start by calculating the  strange scalar matrix elements over the nucleon, assuming
an octet nature of $SU(3)$ symmetry breaking. We follow to ref.\cite{Donoghue}(
see also  the book \cite{Donoghue1} for a review)
in our  calculations \cite{Khriplovich}, but with a small difference in details.
We present these results for completeness of the talk.

The results of the fit to the data on $\pi N$ scattering presented in
\cite{Koch}
lead to the following estimates for the so-called $\sigma$ term
\be
\label{1}
\frac{m_u+m_d}{2}\la p|\bar{u}u+\bar{d}d|p\ra=(64\pm 8 MeV).
\ee
(Here and in what follows we omit kinematical structure like
$\bar{p}p$ in expressions for matrix elements.)
Taking the values of quark masses to be
$m_u=5.1\pm0.9MeV~,m_d=9.3\pm1.4MeV~,m_s=175\pm25MeV$ \cite{Leutwyler},
 from (\ref{1}) we have
\be
\label{2}
\la p|\bar{u}u+\bar{d}d|p\ra\simeq   9.
\ee
Further, assuming octet-type $SU(3)$ breaking to be responsible for the mass splitting in the baryon octet, we find
\be
\label{3}
\la p|\bar{u}u-\bar{d}d|p\ra=\frac{m_{\Xi}-m_{\Sigma}}{m_s}=   0.7,
\ee
\be
\label{4}
\la p|\bar{u}u+\bar{d}d -2\bar{s}s|p\ra=3\frac{m_{\Xi}-m_{\Lambda}}{m_s}=  3.4.
\ee
Here $m_{\Xi},m_{\Sigma},m_{\Lambda}$
are masses of $\Xi,\Sigma,\Lambda$ hyperons respectively.
The values (\ref{3}), (\ref{4}) are quite reasonable: the former is close
 to the difference of the number of $u$ and $d$ quarks in a proton (should be $1$), and the
 latter is close
to the total number of valence quarks
$u$ and $d$ in a nucleon (see below).
from (\ref{2}-\ref{4}) one obtains:
\be
\label{5}
\la p|\bar{u}u |p\ra\simeq   4.8,
\ee
\be
\label{6}
\la p|\bar{d}d |p\ra\simeq   4.1,
\ee
\be
\label{7}
\la p|\bar{s}s |p\ra\simeq   2.8.
\ee
We should mention that the accuracy of these equations is not very
high. For example, the error in the value of the $\sigma$ term
already leads to an error of order of one in each matrix element
discussed above. However, these very simple calculations
explicitly demonstrate that the strange matrix element
is by no means small.

We would like to  rewrite the relations (\ref{5}-\ref{7})
  to separate the vacuum contribution to the
nucleon matrix element
 from the valence contribution.
In order to do so, let us define
\be
\label{8}
\la p|\bar{q}q |p\ra\equiv \la p|\bar{q}q |p\ra_0 +\la p|\bar{q}q |p\ra_1,
\ee
where index $0$ labels a (sea) vacuum contribution and index $1$ a valence
contribution  for a quark $q$. We assume that the vacuum contribution which is
related to the sea quarks is the same for all light quarks $u,d,s$.
Thus, the nonzero magnitude for the strange matrix elements
comes exclusively from the vacuum fluctuations. At the same time,
the matrix elements related to the valence contributions are equal to
\be
\label{9}
\la p|\bar{u}u |p\ra_1\simeq   (4.8-2.8)\simeq 2,
\ee
\be
\label{10}
\la p|\bar{d}d |p\ra_1\simeq   (4.1-2.8) \simeq 1.3.
\ee
These values are in remarkable agreement with the numbers $2$ and $1$, which
one could expect from the naive picture of non-relativistic constituent quark
model.
 In spite of the very rough estimations presented above,
we believe we   convinced a reader that :\\
a) a magnitude of the nucleon matrix element
for $\bar{s}s$ is not small;\\
b) the large magnitude for this matrix element is due to the
nontrivial QCD vacuum structure where vacuum expectation values
of $u,d,s$ quarks are developed and they are almost the same in magnitude:
$\la 0|\bar{d}d |0\ra\simeq\la 0|\bar{u}u |0\ra
\simeq\la 0|\bar{s}s |0\ra$.

Once we realized that the phenomenon under discussion
is related to the nontrivial
vacuum structure, it is clear that the best way to
understand such a phenomenon is  to use some method where QCD vacuum
fluctuations and hadronic properties are strongly interrelated.
We believe, that the most powerful analytical nonperturbative method
which exhibits these features is
the QCD sum rules approach \cite{Shif1},\cite{Shif2}.

In what follows we use  the  QCD sum rule method
in order to relate hadronic matrix elements and vacuum characteristics.
Let me emphasize from the very beginning that  we do not use the QCD sum
rules in the standard way: we do not fit them to extract any information about
lowest resonance (as people usually do in this approach), we do not use any
numerical approximation or implicit assumptions about higher states. Instead,
we concentrate on the qualitative relations between hadronic properties
and QCD vacuum structure. We try to explain in qualitative way
some magnitudes for the  nucleon matrix elements which
may look very unexpected from the naive point of view.
At the same time those matrix elements  can be easily understood    in terms
of the QCD vacuum structure.

We close this section with the formulation of the following question:\\
{\it Q: What is the QCD explanation of the unusual properties
mentioned above?} (in particular, the
large magnitude for the strange nucleon matrix element,
a special role of the scalar and pseudoscalar channels et cetera).
Our answer on this question is: \\
{\it A:
Hadronic matrix elements with $0^{\pm}$ quantum numbers
are singled out because of the special role   they  play in the
QCD vacuum structure.}
The next section switches  this answer from a
qualitative remark into the   quantitative description.
\subsection{Strangeness in the nucleon and vacuum structure}
To study the problem
of calculation $\la N|\bar{s}s|N\ra$ using the QCD -sum rules approach, we
consider the following vacuum correlation function \cite{Khriplovich}:
\be
\label{11}
T(q^2)=\int e^{iqx}dxdy\la 0|T\{\eta(x),\bar{s}s(y), \bar{\eta (0)}\}|0\ra
\ee
at $-q^2\rightarrow\infty$. Here $\eta$ is an arbitrary current
with nucleon quantum numbers. In particular, this current may be chosen in the standard
form $\eta=\epsilon^{abc}\gamma_{\mu}d^a(u^bC\gamma_{\mu}u^c)$. Note, however,
that the results obtained below do not imply such a concretization.
For  the future
convenience we consider the unit matrix kinematical structure in (\ref{11}).

This is the standard first step of any calculation
of such a  kind: Instead of direct calculation of a  matrix element,
we reduce the problem to the computation of some correlation function.
As the next step, we use  the duality and dispersion relations to relate a
physical matrix element to the QCD- based formula for the corresponding
correlation function. This is essentially the basic idea of the QCD sum rules.

In our specific
 case (\ref{11}) due to the absence of the $s$ -quark field in the nucleon
current $\eta$, any substantial contribution to $T(q^2)$ is
connected only with non-perturbative, so-called induced vacuum condensates,
see Fig.1. Such a contribution arises from the region, when
some distances are large: $(y-0)^2\sim (y-x)^2\gg (x-0)^2$.
Thus, it can not be directly calculated in perturbative theory,
instead we code the corresponding large-distance information
in  the  form of a bilocal operator
\be
\label{12}
K=i\int dy\la 0|T\{\bar{s}s(y), \bar{u}u(0)\}|0\ra ,
\ee
see Fig.1
 The similar contributions were considered at first time
in ref.\cite{Balitsky}. For the different applications of this approach
when bilocal operators play essential role, see also
refs.\cite{Ioffe}.

Along  with consideration of the  three-point correlation function (\ref{11}),
we would like to consider the standard two-point  correlator
\be
\label{13}
P(q^2)=\int e^{iqx}dx \la 0|T\{\eta(x), \bar{\eta (0)}\}|0\ra ,
\ee
see Fig.2.
The correlator (\ref{13}) is determined by the  nucleon residues
$\la 0|\eta|N\ra$ and some duality interval $S_0$.
At the same time the correlator (\ref{11})
includes the information on the nucleon matrix element
$\la N|\bar{s}s|N\ra$ also. Comparing (\ref{11})
with (\ref{13}) at $-q^2\rightarrow\infty$, we
arrive to the following relation \cite{Khriplovich}:
\be
\label{14}
\la N|\bar{s}s|N\ra\simeq\frac{-m}{\la\bar{q}q\ra}K ,
\ee
where $m$ is the nucleon mass.
The main assumptions which   have been made in
the derivation of this relation are the
following. First, we made the standard assumption
about local duality for the nucleon. In different words we assumed that
a nucleon saturates both correlation functions
with duality interval $S_0$. The second assumption is that
  the typical scales (or what is the same, duality intervals
in the limit $-q^2\rightarrow\infty$ in the corresponding
    sum rules ( (\ref{11}) and (\ref{13}) )
 are not much different in magnitude from each other.
In this case the dependence on residues $\la 0|\eta|N\ra$ is canceled out
in the ration
and we are left with the matrix element $\la N|\bar{s}s|N\ra$ (\ref{14})
we are interested in.

 Note, that both these assumptions are very likely
to be satisfied  because we know that in most cases
the lowest state (nucleon) does saturate
the sum rules. If it does, than the typical scale
(which   in variety of sum rules  is one and the same and of order of $ 1 GeV^2$)
guarantees that the duality intervals are likely to be very close to each other.
Anyway,  the quantitative analysis of the corresponding sum rules is possible,
however it  is  not our main goal; rather we want  to demonstrate the relation between
matrix elements like $\la N|\bar{s}s|N\ra$ and the corresponding
vacuum properties which are hidden in the correlator $K$ (\ref{12}).
 In principle one could  analyze the sensitivity of the
corresponding QCD sum rules to the lowest state, nucleon. Once it is demonstrated,
we believe that the accuracy of our formula (\ref{14})  is of order $20\%-30\%$
which   is a typical error
for the sum rule approach.

Thus, the calculating of $\la N|\bar{s}s|N\ra$ reduces to
the evaluation of the vacuum correlator
$K$. Fortunately, sufficient information about the latter comes from
the low -energy theorems. We note also that this method of reducing
the nucleon matrix elements to that of the vacuum correlator is directly
generalized to cover the arbitrary scalar $O_S$ or pseudoscalar $O_P$
operator\footnote{We assume of course that these operators
do not contain $u, d$ quarks. Otherwise an additional contribution
which comes from the small distances must be also included.}:
\be
\label{15}
\la N|O_S|N\ra\simeq\frac{-m\bar{N}N}{\la\bar{q}q\ra}
i\int dy\la 0|T\{O_S, \bar{u}u(0)\}|0\ra ,
\ee
\be
\label{16}
\la N|O_P|N\ra\simeq\frac{-m\bar{N}i\gamma_5N}{\la\bar{q}q\ra}
i\int dy\la 0|T\{O_P, \bar{u}i\gamma_5u(0)\}|0\ra .
\ee

The estimation of  the nonperturbative correlator $K$
can be done by using some low-energy theorems.
In this case $K$ is expressed in terms of some vacuum condensates
 \cite{Khriplovich}:
\be
\label{17}
K=i\int dy\la 0|T\{\bar{s}s(y), \bar{u}u(0)\}|0\ra \simeq
\frac{18}{b}\frac{\la\bar{q}q\ra^2}{\la \frac{\alpha_s}{\pi}G_{\mu\nu}^2\ra}
\simeq 0.04 GeV^2 ,
\ee
where
$b=\frac{11}{3}N_c-\frac{2}{3}N_f=9 $ and we use the standard
values for the vacuum condensates\cite{Shif1}:
$$\la \frac{\alpha_s}{\pi}G_{\mu\nu}^2\ra\simeq 1.2 \cdot 10^{-2}GeV^4~~~
\la\bar{q}q\ra\simeq -(250MeV)^3 .$$
With the estimation (\ref{17}) for $K$, our formula (\ref{14}) gives
the following expression for the
nucleon expectation value for $\bar{s}s$
\be
\label{18}
 \la p| \bar{s}s  |p\ra \simeq
-m\cdot\frac{18}{b}\frac{\la\bar{q}q\ra }{\la \frac{\alpha_s}{\pi}G_{\mu\nu}^2\ra}
\simeq 2.4  ,
\ee
which is very close to the naive estimation (\ref{7}).
 Let us stress: we are not pretending to have made
a reliable calculation of the matrix element $ \la p| \bar{s}s  |p\ra$
here. Rather, we wanted to emphasize on the    qualitative picture which
demonstrates the close relation
between nonvalence matrix elements and QCD vacuum structure.

We close this section by noting that
the method presented above gives very simple
 physical explanation of why  the Zweig rule in
  the scalar and pseudoscalar
channels  is badly broken  and at the same time, in the vector channel the
Zweig rule works well. In particular, the matrix element
$\la N|\bar{s}\gamma_{\mu}s|N\ra$ is expected to be very small as well as
  the corresponding coupling constant $g_{\phi NN}$ does.
  In terms of   QCD
such a smallness corresponds to the numerical suppression ($10^{-2}-10^{-3}$)
of the nondiagonal correlation function
$\int dx \la 0|T\{\bar{s}\gamma_{\mu}s(x), \bar{u}\gamma_{\nu}u(0)\}|0\ra$
in comparison with the diagonal one
$\int dx \la 0|T\{\bar{u}\gamma_{\mu}u(x), \bar{u}\gamma_{\nu}u(0)\}|0\ra$,
see QCD-estimation in \cite{Shif1}.
In the scalar and pseudoscalar channels the diagonal and non-diagonal channels
have the same order of magnitude.

In the next few sections we discuss some applications of the obtained results.
\subsection{In the world where  $s$ quark is massless.}
 We would like to look at formula (\ref{17})
from the different side. Namely, we note that $K$ not only enters
expression (\ref{14}), but also determines the variation
of the condensate $\la\bar{u}u \ra$ with $s$ quark mass:
\be
\label{19}
\frac{d}{dm_s}\la\bar{u}u \ra=
-i\int dy\la 0|T\{\bar{s}s(y), \bar{u}u \}|0\ra =-K\simeq -0.04GeV^2.
\ee
To understand how large this number is and in order to make
some rough estimations, we assume that this behavior can be
extrapolated from physical value $m_s\simeq 175 MeV$ till $m_s=0$.
In this case we estimate that
\be
\label{20}
\mid\frac{ \la\bar{u}u \ra_{m_s=175}-\la\bar{u}u \ra_{m_s=0}}
{\la\bar{u}u \ra_{m_s=175}}\mid\simeq 0.5.
\ee
Such a decrease of $\mid\la\bar{u}u \ra\mid$ by a factor of two
as $m_s$ varies from $m_s\simeq 175 MeV$ to $m_s=0$ is a very important
consequence of the previous discussions: Once we accept
the relatively large magnitude for the nucleon matrix element
$\la p| \bar{s}s  |p\ra \simeq 2.4$, we are forced
to accept the relatively large variation of the light quark condensate
as well. This statement is the direct consequence of QCD, see
(\ref{20}).

We note that this result does not seem very surprising
since other vacuum condensates, e.g. $\la \frac{\alpha_s}{\pi}G_{\mu\nu}^2\ra$
possess analogous properties \cite{Novikov}.
From the microscopic point of view, decrease
of absolute values of vacuum matrix elements with the
decrease of the $s$ quark mass is expected since any  topologically
nontrivial vacuum configurations, e.g. instantons, are suppressed
by light quarks. The corresponding numerical calculation
is very difficult to perform, however a  qualitative
picture of the QCD vacuum structure definitely supports this idea
\cite{Shuryak}.
\subsection{$s$ quark and the nucleon mass.}
We would like to discuss here one more fundamental
characteristic of the hadron world: the nucleon mass
and its dependence on the strange quark.
We start our discussion from the following well known
result: the nucleon mass is determined by the trace of the energy
-momentum tensor $\theta_{\mu\mu}$ and in the chiral limit
$m_u=m_d=m_s=0$ the nonzero result comes exclusively
from the strong interacting gluon fields:
\be
\label{21}
m=-\frac{b}{8}\la N| \frac{\alpha_s}{\pi}G_{\mu\nu}^2 |N\ra,~~
m_u=m_d=m_s=0.
\ee
However, as we know, in our world the strange quark is not massless,
but rather it requires some (large enough) mass ($\sim 175 MeV$).
As we have seen (\ref{20}), the
nonzero mass of $s$ quark considerably changes the vacuum properties
of the world. Thus, we would expect that it might have strong influence on
the nucleon mass as well. The main argument  which supports
 this point of view is the same as before, and   is based on
our general philosophy that the nucleon matrix elements and
vacuum properties are tightly related. So, if the
strange quark has strong influence on the vacuum properties than
its impact on the nucleon mass     should also be strong.

In order to check these reasons it would be useful
to calculate the strange quark contribution into the
nucleon mass directly and independently from the gluon contribution (\ref{21}).
Fortunately, it can be easily done by using our previous estimation (\ref{18})
for the nucleon matrix element and exact expression
for the trace of the energy-momentum tenzor with taking into account
nonzero quark masses:
\be
\label{22}
m=+\la N|\sum_{q}m_q\bar{q}q|N\ra-
\frac{b}{8}\la N| \frac{\alpha_s}{\pi}G_{\mu\nu}^2 |N\ra,
\ee
where  sum over $q$ is sum over all light qurks $u, d, s$.
One can easily see from (\ref{1}) that  $u, d$ contribution into the nucleon mass
does not exceed $7\%$; thus we can safely neglect
this. At the same time, adopting the values (\ref{7}),(\ref{18})
for $\la p| \bar{s}s  |p\ra $ and $m_s\simeq 175 MeV$ \cite{Leutwyler},
one can conclude that considerable part of the nucleon mass
(about $45\%$) is due to the strange quark.
In this case the gluon contribution into the nucleon mass
is far away from the chiral $SU(3)$ prediction (\ref{21})
and approximately equals to
\be
\label{22a}
-\frac{b}{8}\la N| \frac{\alpha_s}{\pi}G_{\mu\nu}^2 |N\ra\sim 520 MeV.
\ee

This rough estimation
confirms our argumentation that  a variation of the strange quark mass
from its physical value to zero,  may  considerably change some
vacuum characteristics as well as nucleon matrix elements.

The simple consequence of this result is the observation that
the {\bf quenched approximation} in the  lattice calculations
is not justified simply  because such a calculation
 clearly not  accounting
the fluctuations of the strange (non-valence) quark as
well as vacuum fluctuations
of $u$ and $d$ quarks. As we argued above, the nucleon mass undergoes some
influence from $ s$ quark.

How one can understand these results within the framework
of the QCD sum rules? Let us recall that in the QCD sum rules approach
an information about any dimensional parameter is contained
in the vacuum condensates $\la\bar{u}u \ra ,\la G_{\mu\nu}^2\ra, ...$.
As we discussed previously all these condensates varying with $m_s$
considerably. It is important that this variation
certainly proceeds in the right direction: Absolute values
of condensates decrease with decreasing $m_s$. This leads
to a smaller scale in the sum rules and finally, to the decrease
of all dimensional parameters such as $m$. However, it is difficult
to make any reliable calculations because of a large number of
factors playing an essential role in such a calculation.

   \subsection{ Momentum distribution of the  strangeness in  the nucleon .}
We continue our study on  the role of  the strange quark   in
 nucleon with the following remark. We found out earlier that the matrix element
$\la N |\bar{s}s|N\ra$ is not small; we interpreted this result as a result
of   strong vacuum fluctuations which penetrate
  into the nucleon matrix element. Now, we would like to ask the following question:
What is the   mean value of the  momentum
( denoted as $\la k_{\perp}^2\ra_s$ ) of the  $s$ quark inside of a nucleon?
Let us note that this question is not  a pure academic one.
Rather,   the answer on the question might be important for the construction
of a  more sophisticated quark model which
would incorporate the strange context into the  nucleon
wave function.

First of all, let us try to formulate this question in terms of QCD.
We {\it define} the mean    value
$\la k_{\perp}^2\ra_s$
of the momentum  carried by the  strange quark in a nucleon
by the following matrix element:
\be
\label{23}
\la k_{\perp}^2\ra_s \la N| \bar{s}s|N\ra\equiv\la |N \bar{s}
(i\stackrel{\rightarrow}{D_{\perp}})^2s|N\ra ,
\ee
where $i\stackrel{\rightarrow}{D_{\mu}}\equiv i
\stackrel{\rightarrow}{\partial_{\mu}}+gA_{\mu}^a\frac{\lambda^a}{2}$
is the covariant derivative and $A_{\mu}^a$ is gluon field. The arrow shows
 the  quark whose momentum is under discussion.

We assume that the nucleon to be moving rapidly in the $z$-direction.
We are interested in  the momentum distribution in the direction  which is perpendicular
to its motion. Precisely this characteristic has a
dynamical origin. Indeed, as we shall see in a moment,
while we are studying  a nucleon matrix element $\la k_{\perp}^2\ra_s$,
we are actually probing the QCD vacuum properties!
The nucleon motion as  a whole system with arbitrary velocity
does not affect this characteristic.
  Thus, essentially, what
 we discuss is  the, so-called, light cone wave function.
Apart of the reasons mentioned above there   are few more motives to do so:
First of all,
 the light cone wave function ($wf$) with a minimal number
of constituents is a good starting point.
As is known such a function gives the  parametrically leading contributions
to hard exclusive processes. Higher Fock states are  also well defined
in this approach  and
can   be considered separately.
The second reason to work with a light cone   wave function
is there
 existence of the nice relation between that $wf$ and structure function measured
in the  deep-inelastic scattering.
We refer to the review paper \cite{Brod1} for the    introduction
into the subject. The relation to the standard
quark model wave functions
 (see e.g.\cite{Isgur1}) is also worked out.
The relevant discussions can   be found in ref.\cite{BHL}.
Besides of these, we have one more reason to  work with  the light cone $wf$:
  we believe that   this    is the  direction
 where a  valence quark model can be understood
and formulated in    the QCD -terms\cite{Zhit}.

Anyhow,     the formula (\ref{23})
  with the derivatives taken in  the  direction perpendicular to the  nucleon momentum
$p_{\mu}=(E,0_{\perp},p_z)$,
is very natural definition for the mean square of the quark transverse momentum.
  Of course it
is different from the naive, gauge dependent  definition like
$\la N|\bar{s}   \partial_{\perp}^2 s|N\ra $,
because the physical transverse gluon is participant
of this definition.
However, the expression (\ref{23}) is the only possible
way to define the  $\la k_{\perp}^2\ra_s$
in the gauge theory like QCD.
 We believe that such definition is the useful generalization
of the transverse momentum conception for the interactive quark
system.
Let us note that the Lorentz transformation in $z$
direction does not affect the transverse directions. Thus, the
 transverse momentum $\la k_{\perp}^2\ra_s$ calculated from eq. (\ref{23})
  remains unchanged while  we passing from the light cone system to the rest frame system
where a quark model  suppose to be formulated.

Now, let us come back to our definition (\ref{23}) for $\la k_{\perp}^2\ra_s$.
In order to calculate this matrix element,
we use the same trick as before:
we reduce our original problem of the calculating of a nucleon matrix element
to the problem of a computing of the corresponding vacuum correlation function
(\ref{15}):
\be
\label{24}
  \la N |\bar{s}(i\stackrel{\rightarrow}{D_{\perp}})^2s|N\ra
\simeq\frac{-m\bar{N}N}{\la\bar{q}q\ra}
i\int dy\la 0|T\{\bar{s}(i\stackrel{\rightarrow}{D_{\perp}})^2s,
 \bar{u}u(0)\}|0\ra .
\ee
To estimate the right hand side of the eq.(\ref{24})
we introduce an auxiliary vacuum correlation function
\be
\label{25}
i\int dy\la 0|T\{\bar{s}(i\stackrel{\rightarrow}{D_{\mu}}
i\stackrel{\rightarrow}{D_{\nu}})s,
 \bar{u}u(0)\}|0\ra =Cg_{\mu\nu},
\ee
where $C$ is constant.
From the definition (\ref{24}) it is clear that  the
 correlator   we are interested in can be expressed in terms of the
constant $C$:
\be
\label{26}
i\int dy\la 0|T\{\bar{s}(i\stackrel{\rightarrow}{D_{\perp}})^2s,
 \bar{u}u(0)\}|0\ra =-2C.
\ee
At the same time the constant $C$  is given by the
correlation function which contains $G_{\mu\nu}^a $ and not a covariant derivative
$D_{\mu}$:
\be
\label{27}
C=\frac{1}{4}i\int dy\la 0|T\{\bar{s}(i\stackrel{\rightarrow}{D_{\mu}}
i\stackrel{\rightarrow}{D_{\mu}})s,
 \bar{u}u(0)\}|0\ra = \nonumber
\ee
\be
\frac{-1}{8}i\int dy\la 0|T\{\bar{s} igG_{\mu\nu}^a
\frac{\lambda^a}{2}\sigma_{\mu\nu}s,
 \bar{u}u(0)\}|0\ra,
\ee
where we have used the equation of motion and identity\footnote{We
neglect the term proportional to $m_s^2$ in the eq.(\ref{27}).
  It can be  justified by using the estimation  (\ref{17}).}:
\be
\label{28}
D_{\mu}D_{\nu}g_{\mu\nu}s=\gamma_{\mu}\gamma_{\nu}D_{\mu}D_{\nu}s-\sigma_{\mu\nu}
\frac{1}{2}[D_{\mu},D_{\nu}]s=-m_s^2s+
\frac{ig}{2}\sigma_{\mu\nu}G_{\mu\nu}^a\frac{\lambda^a}{2}s
\ee
Now we can estimate the unknown vacuum correlator (\ref{27})
exactly in the same way as we have done before
for the correlation function $K$, see eq.(\ref{17}).
Collecting all formulae (\ref{23}-\ref{28}) together, we arrive to the
following final result for the mean value
of the momentum  carried by the  strange quark in a nucleon:
\be
\label{29}
\la k_{\perp}^2\ra_s \equiv \frac{\la N| \bar{s}
(i\stackrel{\rightarrow}{D_{\perp}})^2s|N\ra }
{\la N| \bar{s}s|N\ra}\simeq\frac{\la N|\bar{s}ig\sigma_{\mu\nu}G_{\mu\nu}^a
\frac{\lambda^a}{2}s|N\ra}
{4\la N|\bar{s}s|N\ra}   \nonumber
\ee
\be
\simeq\frac{\la\bar{s}ig\sigma_{\mu\nu}G_{\mu\nu}^a
\frac{\lambda^a}{2}s\ra}
{4\la\bar{s}s\ra}\cdot\frac{d^{\bar{s}ig\sigma_{\mu\nu}G_{\mu\nu}^a
\frac{\lambda^a}{2}s}}{
d^{\bar{s}s}}
\simeq\frac{1}{4}(0.8GeV^2)\frac{5}{3}\sim 0.33 GeV^2,
\ee
where $d^{O}$ denotes the  dimension of the operator $O$.
For numerical estimation we use the standard
magnitude for the   mixed vacuum condensate
$\la\bar{s}ig\sigma_{\mu\nu}G_{\mu\nu}^a
\frac{\lambda^a}{2}s\ra=0.8GeV^2 \la\bar{s}s\ra$.
The obtained numerical value (\ref{29}) for $\la k_{\perp}^2\ra_s$
looks very reasonable from the phenomenological point of view.

We close this section with a few remarks.
First, the nonvalence  nucleon matrix elements can be expressed in terms of
vacuum condensates in a very nice way. All numerical results obtained in such a way
look very reasonable. As the second   remark,
we emphasize that a study of nonvalence nucleon matrix elements and
an analysis of the QCD vacuum structure is one and the same problem.
We would like to note also, that the nucleon matrix element (\ref{29})
might be very important in the analysis of neutron dipole
moment. This observation is based on  the fact that the so called
chromoelectric dipole moment of the $s$ quark, related to the operator
$\bar{s}g\gamma_5\sigma_{\mu\nu}G_{\mu\nu}^a
\frac{\lambda^a}{2}s $, in many
models    gets  a large factor $\sim m_s/m_q\sim 20$
in comparison with a similar $d$ quark contribution
\cite{He}, \cite{Khriplovich}, \cite{Khriplovich1}.
At the same time, as we can see from (\ref{29})
there is no any suppression due to the presence of the $s$ quark
in the corresponding nucleon matrix elements.

\section{Strangeness in the nucleon, $0^-$channel.}
\subsection{Singlet axial constant $g_{A}^{0}$.}
In this section we   discuss the
contribution of the strange quarks into the
nucleon matrix elements similar to eq. (\ref{22}),
with  the only difference that we switch the scalar channel $\bar{s}s$ into the  pseudoscalar one
$\bar{s}i\gamma_5s$. In our previous study of the scalar channel we
  concluded that the considerable part of the nucleon mass
(about $40\%$) is due to the strange quark\footnote{In the chiral limit
$m_s\rightarrow 0$, the corresponding contribution is zero, of course.}. We made this estimation
by using two following facts: First, we knew the mass of the nucleon (left hand side
of the eq. (\ref{22})), which is considered as an experimental  data.
The second, we calculated independently the matrix element $\la N| \bar{s}s|N\ra$.
Comparing this theoretical result (\ref{18}) with (\ref{22}),
we have made aforementioned conclusion about a serious deviation from the chiral $SU(3)$ limit.

We want to repeat all these steps for the pseudoscalar channel also.
In this case the equation analogous to (\ref{22}) looks as follows:
\be
\label{30}
2mg_{A}^{0}\bar{p}i\gamma_5 p=+\la N|\sum_{q}2m_q\bar{q}i\gamma_5q|N\ra+
\frac{3}{4}\la N| \frac{\alpha_s}{\pi}G_{\mu\nu}\tilde{G_{\mu\nu}} |N\ra,
\ee
where  sum over $q$ is the sum over all light qurks $u, d, s$
and $g_{A}^{0}$ is the nucleon axial constant in the  flavor singlet channel.
The world average is: $g_{A}^{0}=0.27\pm 0.04$, \cite{Ellis}.

Now, we would like to repeat all steps which would bring us to the
conclusion similar to eq.(\ref{22}) for the pseudoscalar case. We shall try
to answer on the following question: what is the  strange quark contribution
in  the formula (\ref{30})? Let us recall that in the chiral limit
$m_u=m_d=m_s=0$ the nonzero contribution comes exclusively from the gluon term
in the close analogy with formula (\ref{21}):
\be
\label{31}
2mg_{A}^{0}\bar{p}i\gamma_5 p=
\frac{3}{4}\la N| \frac{\alpha_s}{\pi}G_{\mu\nu}\tilde{G_{\mu\nu}} |N\ra,
~~~~~~~m_u=m_d=m_s=0.
\ee
Thus, in order to answer on the question formulated above,
we have to estimate the matrix element
\be
\label{32}
\la p| 2m_s\bar{s}i\gamma_5s|p\ra
\ee
in somewhat independent way \footnote{We neglect
$u,d$ qurk contributions into the formula (\ref{30}) by the obvious reasons.}.
First of all, the relevant contribution
with octet quantum numbers   ($\eta$)  can be easily evaluated by the  standard
technics. One should  take the derivative from
the octet, anomaly- free, current $\sim \bar{u}\gamma_{\mu}\gamma_5u+
\bar{d}\gamma_{\mu}\gamma_5d-2\bar{s}\gamma_{\mu}\gamma_5s$.
The result is:
\be
\label{33}
\la p| 2m_s\bar{s}i\gamma_5s|p\ra_{\eta}=
-m(3F-D) \bar{p}i\gamma_5p ,
\ee
where $D\simeq 0.63 g_A$ and $F\simeq 0.37 g_A$ are the standard $SU(3)$
parameters.
One could expect that
the similar contribution with
singlet quantum numbers ($\eta'$)  is also large, although
it is zero in the chiral limit where $m_s=0$.

We shall estimate the corresponding  contribution
with $\eta'$- quantum numbers by
using our previous trick (\ref{16}). Namely, we reduce our original
problem of calculation of the nucleon matrix element to the problem of the
computation of certain vacuum correlation function
where we should limit ourself by calculating the contribution with singlet
quantum numbers only:
\be
\label{34}
\la p| 2m_s\bar{s}i\gamma_5s|p\ra  \simeq
\frac{-m\bar{p}i\gamma_5p}{\la\bar{q}q\ra}
i\int dy\la 0|T\{2m_s\bar{s}i\gamma_5s, \bar{u}i\gamma_5u(0)\}|0\ra .
\ee
 In order to make the corresponding estimations we need to know
the following $\eta'$- matrix elements:
$\la 0|\bar{s}i\gamma_5s|\eta'\ra$ and $\la 0|\bar{u}i\gamma_5u|\eta'\ra$.
The PCAC does not provide us with the corresponding information, however a
quark model prejudice suggests that
\be
\label{35}
\la 0|\frac{1}{\sqrt{2}}(\bar{u}i\gamma_5u-\bar{d}i\gamma_5d)|\pi\ra\simeq
\la 0|\frac{1}{\sqrt{6}}(\bar{u}i\gamma_5u+\bar{d}i\gamma_5d-2\bar{s}i\gamma_5s)|
\eta\ra\simeq       \nonumber
\ee
\be
\la 0|\bar{u}i\gamma_5s|K\ra\simeq
\la 0|\frac{1}{\sqrt{3}}(\bar{u}i\gamma_5u+\bar{d}i\gamma_5d+\bar{s}i\gamma_5s)|
\eta'\ra\simeq -\frac{2\la\bar{q}q\ra}{f_{\pi}}
\ee
The strong support in favor  that the relations (\ref{35}) are to be correct,
comes from the analysis of the two photon decays of $\pi,\eta,\eta'$,
see e.g.\cite{Donoghue1}.
All of these decay amplitudes have the same Lorentz structure
and determined by the matrix elements (\ref{35}), therefore, the quark model
prediction is found to work surprisingly well in this particular case.
Combine  the formulae (\ref{33}-\ref{35}) we arrive to the
following estimation:
\be
\label{36}
\la p| 2m_s\bar{s}i\gamma_5s|p\ra =
2m\bar{p}i\gamma_5p \{-\frac{1}{2}(3F-D)- \frac{4m_s\la\bar{q}q\ra}{3f_{\pi}^2m_{\eta'}^2}\}
\nonumber
\ee
\be
\simeq (-0.3+0.16)2m\bar{p}i\gamma_5p\simeq (-0.14)2m\bar{p}i\gamma_5p,
\ee
where we used $3F-D\simeq 0.6$ for the numerical estimation.
The two terms in this formula  are the octet   and
  singlet  contributions correspondingly. One should note, that in spite of the fact
that the singlet term is parametrically
suppressed in the limit $m_s=0$,   this
  contribution numerically is not small. It is only by a factor of two less than the
parametrically leading term.

Now, let us come back to eq.(\ref{30}).
We would like to answer on the previously formulated question:
what is the $s$ quark contribution into the formula (\ref{30})?
From our estimation (\ref{36}) we suggest  the following pattern
of saturation of the experimental data for $g_A^0=0.27\pm 0.04$:
\be
\label{37}
2m \bar{p}i\gamma_5 p (0.27\pm 0.04)=+\la p| 2m_s\bar{s}i\gamma_5s|p\ra+
\frac{3}{4}\la p| \frac{\alpha_s}{\pi}G_{\mu\nu}\tilde{G_{\mu\nu}} |p\ra
\nonumber
\ee
\be
\simeq (-0.14)2m \bar{p}i\gamma_5 p +(+0.41)2m \bar{p}i\gamma_5 p.
 \ee
Here the first term is due to the strange quark contribution
from (\ref{36}) and the second one
is due to the gluon contribution. We assign an average
number $0.41$ for the gluon contribution in order
to match   the experimental data for $g_A^0$.

Two remarks are in order. First,
The strange-quark   and the gluon terms contribute with the
opposite signs into $g_A^0$. In the   formula
for mass (\ref{22}) the similar terms interfere constructively, with the same signs.
It is very easy to understand the difference:
in the pseudoscalar channel we have the Goldstone boson, $\eta$, whose
total contribution is zero into the sum (\ref{30}) because
of the octet origin of the $\eta$ meson.
However, the $\eta$ meson contributions into the
matrix elements $\la p| 2m_s\bar{s}i\gamma_5s|p\ra$
and into the gluon operator,
  taken separately, are not zero.
 Moreover, its contribution to the $\la p| 2m_s\bar{s}i\gamma_5s|p\ra$
has the opposite sign   to
the  $\eta'$ contribution (because of the difference in the
quark context, see (\ref{35})). Even more, it has a  parametrical enhancement.
We have nothing like that in the scalar channel (\ref{22}),
where the flavor -singlet states  dominate.

The second remark is the observation that,
like in the scalar channel, the
  strange quark operator gives a
noticeable contribution into the final formula (\ref{37})
in spite of the fact that
in the chiral limit the corresponding contribution
is   zero as we mentioned earlier (\ref{31}).
\subsection{Singlet axial constant of heavy particles.}
In this section we   try to answer on the following question:
How one can check  the
  estimation (\ref{37})
about noticeable contribution of the
strange quark? More specific question we would like to
know: Is it possible to measure a nucleon matrix element where
some independent
combination of  those operators
enters?
If   answer were ``yes'', we would be able to
find the contribution of each
 term  separately.

To answer on this question we suggest to consider the weak neutral
current containing an isoscalar axial component associated
with nonvalence quarks \cite{Collins}:
\be
\label{38}
\la p| \bar{c}\gamma_{\mu}\gamma_5c- \bar{s}\gamma_{\mu}\gamma_5s
+\bar{t}\gamma_{\mu}\gamma_5t- \bar{b}\gamma_{\mu}\gamma_5b|p\ra
\equiv g_A^{heavy}\bar{p}\gamma_{\mu}\gamma_5p
\ee
Before to go into details, let us mention, that
on the quantum level the current divergence of the massive quark field
has the following form:
\be
\label{39}
\partial_{\mu}\bar{Q}\gamma_{\mu}\gamma_5Q=2m_Q\bar{Q} i\gamma_5Q+
 \frac{\alpha_s}{4\pi}G_{\mu\nu}\tilde{G_{\mu\nu}},
\ee
where the first term is the standard one and the second term is
due to the anomaly. There are many ways to understand the
origin of the anomaly; basically it arises from the necessity
of the ultraviolet regularization of the theory.
In the heavy quark mass limit, one can expand $
2m_Q\bar{Q} i\gamma_5Q$ term in the eq.(\ref{39}) with the following result
\cite{Shif1},\cite{Shif2}:
\be
\label{40}
2m_Q\bar{Q} i\gamma_5Q|_{m_Q\rightarrow\infty}=
 -\frac{\alpha_s}{4\pi}G_{\mu\nu}\tilde{G_{\mu\nu}}+
c\frac{G\tilde{G}G}{m_Q^2}+0(\frac{1}{m_Q^4})+...
\ee
where all coefficients, in principle, can be calculated.
We are interested, however,  in the
leading term $\sim G_{\mu\nu}\tilde{G_{\mu\nu}}$ only.

One can easily note that the leading term in the expansion (\ref{40})
 has the same structure as
an anomaly term (\ref{39}) and it goes with the opposite sign\footnote{
The opposite signs of those contributions can be easily
understood in terms of the
Pauli-Villars regulator fields with mass $M_{PV}\rightarrow\infty$.
As is known these fields are introduced into the theory for the
regularization purposes and they play crucial role in the  calculation of the
anomaly (\ref{39}). Regulator contribution is obtained, by definition,
   by  a replacement $m_Q\rightarrow M_{PV}$ in
the corresponding formula. It goes, by definition, with relative sign minus.
  From such a calculation it is clear
that the leading terms which do not depend on mass, are
canceled out in a full agreement with an explicit formula (\ref{39},\ref{40}).}.
 Thus, the terms, which do not depend on mass
are canceled  out   and
we left with a term $\sim\frac{G\tilde{G}G}{m_Q^2}$ which vanishes
in the limit $m_Q\rightarrow\infty$. Such a vanishing
of the heavy quark contribution into the nucleon matrix element is
 in a perfect agreement
with a physical intuition that a  nucleon does not contain
any heavy quark fields, at least in the limit $m_Q\rightarrow\infty$.

The situation with strange $s$ quark in the formula (\ref{38})
is much more complicated. This quark is not heavy enough to
apply the arguments given above. Thus, we should keep
all operators in the formula (\ref{39}) for the current divergence
in the original form:
\be
\label{41}
\la p|2m_s\bar{s} i\gamma_5s+
 \frac{\alpha_s}{4\pi}G_{\mu\nu}\tilde{G_{\mu\nu}}|p\ra=
-g_A^{heavy}2m\bar{p}i\gamma_5p,
\ee
where we have neglected the terms $\sim \frac{1}{m_Q^2}$
for $c, b,t $ quarks in according with our previous discussion\footnote{
The corresponding estimations even for the
lightest heavy $c$-quark support this viewpoint, see the next section.}.

As we already mentioned, the measurement of the
constant $g_A^{heavy}$ (\ref{41}) gives an independent information
  complementary to the singlet axial constant measurement
$g_A^0$, (\ref{37}). If we knew those constants with high
enough precision, we would be able to find out  both
  nucleon matrix elements:$\la p|2m_s\bar{s} i\gamma_5s|p\ra$
and  $\la p|\frac{\alpha_s}{4\pi}G_{\mu\nu}\tilde{G_{\mu\nu}}|p\ra$.
At the moment the experimental errors for
$g_A^{heavy}=0.15\pm0.09$, \cite{Ahrens} are large:
  the result is only two standard deviations from zero.

The best we can do at the moment is to estimate $g_A^{heavy}$
from our previous calculations (\ref{37}).
If we literally take the values $-0.14$ and $0.41$ from the formula (\ref{37}),
we   get the result for $g_A^{heavy}$ which is compatible with zero:
\be
\label{42}
-g_A^{heavy}2m\bar{p}i\gamma_5p=
\la p|2m_s\bar{s} i\gamma_5s+
 \frac{\alpha_s}{4\pi}G_{\mu\nu}\tilde{G_{\mu\nu}}|p\ra \simeq \nonumber
\ee
\be
 \simeq (-0.14+\frac{1}{3}0.41)2m\bar{p}i\gamma_5p,~~g_A^{heavy}\simeq 0.
\ee
From our point of view this is an interesting observation which essentially
says that the strange quark operator {\bf together with its anomalous
part} gives nearly {\bf vanishing }
contribution into the nucleon matrix
element. As we discussed earlier, this is certainly true
for any heavy quark. What is surprised us, that
estimation (\ref{42}) apperently says that this is   true
even for $s$ quark (which
is by no means can  be considered as a heavy quark).
If we accept this point, we should interpret
a nonzero magnitude of  $g_A^0$ (\ref{37}) as a  contribution
coming exclusively from  the light $u,d$ quarks and their anomalous  parts.
 As we mentioned, the $s$ quark term together with its anomalous part
 gives almost vanishing contribution into eq.(\ref{37}).
Such an interpretation is in a very good agreement with the
valence quark model philosophy, where the $s$ quark
does not play any essential role.

Let us note that this interpretation is very different
from the old simplest assumption on the spin of the strange quark
in the nucleon, see e.g.\cite{Ellis},\cite{Donoghue1}.
In our interpretation we understand the strange quark
contribution as a joined contribution of $s$ field as well as   its regulator
field (or what is the same, its anomalous contribution).
\section{The charmed quark in the nucleon.}
In this section we would like to extend our analysis for the
$c$ quark. The reason  to do so is twofold: First,
       the $c$ quark
   is a heavy enough to use the standard
 $1/m_c$ expansion similar to (\ref{40}). Secondly,
 the charmed quark is light enough
to get a reasonably  large   effect from this expansion.

We start from the pseudoscalar channel
and keep only the first term in the heavy quark expansion (\ref{40}):
\be
\label{43}
 \la p|\bar{c} i\gamma_5c|p\ra \simeq
 -\la p|\frac{\alpha_s}{8m_c\pi}G_{\mu\nu}\tilde{G_{\mu\nu}} |p\ra
\sim \frac{-0.41}{6m_c}2m\bar{p}i\gamma_5p\sim -0.1\bar{p}i\gamma_5p
\ee
where, for the numerical estimate,
we use
 the value (\ref{37}) for the gluon matrix element
over nucleon,
and $m_c\simeq 1.3 GeV$
for the charmed quark mass\footnote{
Let us note, that this value for mass corresponds to
 the high enough normalization point of order of $m_c$. In principle, one should
renormalize this value to the low normalization point. We neglect this
small logarithmic effect in this paper.}.
This value should be compared with the similar matrix element
of the strange quark over nucleon:
\be
\label{44}
 \la p|\bar{s} i\gamma_5s|p\ra \sim
\frac{-0.14}{2m_s}2m\bar{p}i\gamma_5p\sim -0.8\bar{p}i\gamma_5p,
\ee
where we use     formula (\ref{36}) for the numerical
estimation of the matrix element $\la p|\bar{s} i\gamma_5s|p\ra$.
The ratio of these values are in remarkable agreement with the
ratio of their mass: $\frac{m_c}{m_s}\sim\frac{1.3GeV}{0.175GeV}\sim
7.5$.

Our next example is the scalar matrix element. In this case
one can use the heavy quark expansion similar to formula
(\ref{40}), but for the scalar channel\cite{Shif2}:
\be
\label{45}
 \la p|\bar{c}  c|p\ra \simeq
 -\la p|\frac{\alpha_s}{12m_c\pi}G_{\mu\nu}^a G_{\mu\nu}^a |p\ra
\sim \frac{2\cdot 520MeV}{27m_c}  \sim 0.03.
\ee
For the numerical estimation in this formula  we adopted
the value (\ref{22a}) for the gluon matrix element
over nucleon. The magnitude (\ref{45}) for the
charmed quark is approximately hundred times less than the
corresponding matrix element for the strange quark (\ref{18}):
\be
\label{46}
 \la p| \bar{s}s  |p\ra  \simeq 2.4  .
\ee
 This is
in a big contrast with pseudoscalar channel, where the
corresponding ratio was about a  factor   ten larger.

We conclude this section with few   remarks.
First of all, the matrix elements (\ref{43}),(\ref{45}) for the
charmed quark are expressed in terms of the gluon operators.
For the heavy quark this is exact consequence of the QCD.
Corrections to these formulae can be easily  estimated.
One can show that  they are
   small
for the $c$ quark. The problem of the evaluation of the
gluon matrix elements over  nucleon is the different problem.
However, we believe that   from  the measurements of
 $g_A^0$ (\ref{30}) and  from $\pi N$ scattering  (\ref{1})  we
know those matrix elements with a reasonable accuracy.
Thus, we expect the same accuracy for the  matrix elements
$\la p|\bar{c}  c|p\ra$ and $\la p|\bar{c} i\gamma_5 c|p\ra$.

Our next remark is the observation that results
(\ref{43}), (\ref{45}) essentially give
a normalization for the intrinsic charm quark
 component in the proton.
 This is very important characteristic of the nucleon.
It might play an essential role in the explanation
 of a discrepancy between charm hadroproduction and perturbative QCD
calculations. We refer to the original paper \cite{Brodsky}
\footnote{See also recent paper \cite{Ingelman}.}
on this subject where the hypothesis of intrinsic charm
quarks in the proton was introduced.
The experimental fit
in the framework of this paper\cite{Brodsky}
suggests that the probability to have an intrinsic
charm in the proton is about $\sim 0.3\%$\cite{Hoff} and
$(0.86\pm0.60)\%$\cite{Harris}.
These numbers can not be  related directly
to the   matrix element  (\ref{45}) we calculated.
However, they give some general scale of this phenomenon.
We hope that in future, some   more sophisticated, QCD-based methods,
will lead us to    deeper understanding of the effects related to
  intrinsic charm component in the nucleon.
 \section{Conclusion}
We believe that the main result of the present analysis
is the observation that   non-valence quarks play
an important role in the physics of nucleon.
 However we should stress that such an interpretation
does not contradict to the   bag model \cite{Donoghue1}, where
the nucleon matrix element
\be
\label{47}
\la N|\bar{s}s|N\ra\simeq -{\la\bar{s}s\ra}V
\ee
is related to  some vacuum characteristics
(like condensate or volume of the bag $V$) of the model.
It is clear that the chiral vacuum condensate of the strange quark is large.
So, there is no reason to expect that the corresponding nucleon
matrix element is small. The same argument can be applied
for the arbitrary mixed vacuum condensates also (they are
presumably   not zero\cite{Zhit}). This information can be translated,
in according to (\ref{29}), into the knowledge about transverse
momentum distribution of the strange quark in a nucleon.

In our approach this relation
$ vacuum\Longleftrightarrow nucleon$ is clearly seen.
Thus,   by studying  the vacuum properties of the QCD, we
essentially
 study
some interesting nucleon matrix elements which can be
experimentally measured.

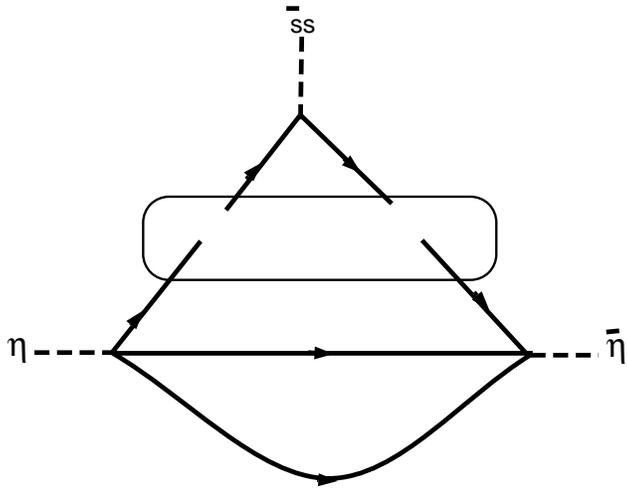
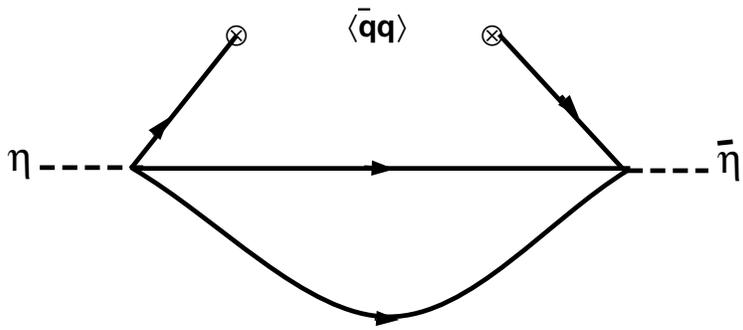
\end{document}